# Authenticity in Authorship: The Writer's Integrity Framework for Verifying Human-Generated Text

Sanad Aburass, Maha Abu Rumman

Department of Computer Science

Maharishi International University

Fairfield, Iowa, USA

saburass@miu.edu, maha.rumman@miu.edu

**Abstract**

The "Writer's Integrity" framework introduces a paradigm shift in maintaining the sanctity of human-generated text in the realms of academia, research, and publishing. This innovative system circumvents the shortcomings of current AI detection tools by monitoring the writing process, rather than the product, capturing the distinct behavioral footprint of human authorship. Here, we offer a comprehensive examination of the framework, its development, and empirical results. We highlight its potential in revolutionizing the validation of human intellectual work, emphasizing its role in upholding academic integrity and intellectual property rights in the face of sophisticated AI models capable of emulating human-like text. This paper also discusses the implementation considerations, addressing potential user concerns regarding ease of use and privacy, and outlines a business model for tech companies to monetize the framework effectively. Through licensing, partnerships, and subscriptions, companies can cater to universities, publishers, and independent writers, ensuring the preservation of original thought and effort in written content. This framework is open source and available here, https://github.com/sanadv/Integrity.github.io

## I. Introduction

In the rapidly evolving landscape of digital content creation, the distinction between human-generated and artificial intelligence (AI)-generated text has become a focal point of ethical, academic, and intellectual debate (Barrett & Pack, 2023; Hall, 2024). With the advent of sophisticated generative AI models, the ability to produce text that mimics human-like quality and creativity has raised significant concerns within academic institutions, research journals, and the broader publishing community (Walters, 2023; Walters & Wilder, 2023). These concerns are primarily centered around the authenticity and integrity of works required for academic degrees, research dissemination, and literary publications, where the value is traditionally placed on the human intellect and effort (Chan & Hu, 2023; Farrelly & Baker, 2023). Recognizing the challenges in distinguishing between human and AI-authored texts, we introduce a novel framework, "Writer's Integrity," designed to ensure the authenticity of human-generated content. The "Writer's Integrity" framework stems from the premise that the process of writing, inherent to human authors, is fundamentally different from the capabilities of AI. Human writing is characterized by its iterative nature, including drafting, revising, and editing, often accompanied by a range of typing speeds and error correction patterns. In contrast, AI models can generate polished, sophisticated text instantaneously, without the need for such revisions (Kaliyar, 2020; Kalyan et al., 2021). Current tools aimed at detecting AI-generated text often fall short, producing false positives and negatives due to their reliance on textual analysis alone. These tools struggle to differentiate effectively between human and AI authors, especially in cases where humans use paraphrasing tools or when AI manages to mimic human writing patterns closely (Akram, 2023). To address these limitations, the "Writer's Integrity" framework proposes a novel approach that does not analyze the text itself but rather the process and behavior behind its creation. This includes monitoring metrics such as typing speed, frequency of edits, and the ratio of pasted to original text. By focusing on these aspects, our framework aims to capture the uniquely human elements of the writing process, thereby providing a more reliable method for distinguishing between human and AI authors. Moreover, the framework addresses concerns related to the use of paraphrasing tools, recognizing them as legitimate aids in the human writing process, provided the original ideas and efforts remain human. A cornerstone of this framework is the issuance of a "Writer's Integrity Certificate." This certificate serves as a tangible attestation to the human authorship of a text, based on the analysis of the writing process. It provides a detailed log of the writing journey, including typing speed, revisions, and the originality of content, which can be shared with academic supervisors, journal reviewers, or publishers. This certificate is not just a document but a testament to the human effort, creativity, and intellectual engagement that went into the creation of the work. It stands as a bulwark against the encroachment of AI on spaces where human intellectual contribution is of paramount importance. This paper delves into the development and operational principles of the "Writer's Integrity" framework, highlighting its potential to revolutionize the validation of human intellectual work. By focusing on the certification process, we underscore the framework's utility in academic integrity, the affirmation of human creativity, and the ethical augmentation of human work with AI.

## II. Literature Review on the Detection of AI-Generated Text

The detection of AI-generated text, particularly text produced by models like ChatGPT, poses significant challenges for current detection tools. Recent studies have attempted to assess the effectiveness of these tools, revealing a complex landscape of accuracy, biases, and limitations. A comprehensive study by Weber-Wulff et al. (2023) evaluated the general functionality of 12 publicly available tools and two commercial systems (Turnitin and PlagiarismCheck), revealing a significant bias towards classifying outputs as human-written rather than detecting AI-generated text. The study highlighted that these tools are neither accurate nor reliable, with content obfuscation techniques further deteriorating their performance (Weber-Wulff et al., 2023). Elkhatat et al., focused on the diagnostic accuracy of AI content detectors, employing a normalization process to standardize results across different tools. It was found that the tools often misclassified AI-generated text as human-written (false negatives) and vice versa (false positives), raising concerns about their effectiveness in accurately identifying AI-generated content (Elkhatat et al., 2023). Zhang et al. (2023) demonstrated that large language models (LLMs) can be guided to evade AI-generated text detection through substitution-based optimization strategies. This involves making semantic substitutions at both word and sentence levels to minimize the predicted probability of text being recognized as AI-generated. The study underscores the potential for LLMs to circumvent detection, questioning the efficacy of current detection tools against adaptive and sophisticated AI-generated texts (Lu et al., 2023). The effectiveness of detection tools varies across different AI models and tasks. For instance, some tools may perform better at identifying content generated by earlier versions of AI models but struggle with content from more advanced versions. This inconsistency can lead to varying levels of detection accuracy, complicating efforts to maintain academic integrity and combat misinformation (Bellini et al., 2024). The limitations of AI-generated text detection tools have significant implications for academic integrity and the fight against misinformation. The inability of these tools to reliably distinguish between human and AI-generated texts could undermine efforts to ensure the authenticity and credibility of information across various domains (Chaka, 2024).

In summary, the literature suggests that while AI-generated text detection tools represent a crucial line of defense against the misuse of AI in content creation, they currently face significant challenges in terms of accuracy, reliability, and resistance to evasion techniques. These findings call for ongoing research and development to enhance the capabilities of detection tools and ensure they can effectively meet the challenges posed by advancing AI technologies.

In response to the limitations of textual analysis-based AI detection tools, the "Writer's Integrity" framework proposes a novel approach that shifts the focus from the product of writing to the process. This framework assesses human authorship by analyzing behavioral metrics such as typing speed, frequency of edits, and the ratio of pasted versus original text. Central to this framework is the issuance of a "Writer's Integrity Certificate," which serves as evidence of human authorship, grounded in the processual and iterative nature of human writing. This innovative approach addresses the gaps left by existing tools, providing a robust mechanism for authenticating human-generated content.

The proliferation of AI-generated content raises profound questions about academic and intellectual integrity. The ease with which AI can generate sophisticated text challenges traditional concepts of authorship and originality, necessitating new measures to safeguard the authenticity of human intellectual efforts. The "Writer's Integrity" framework emerges as a critical tool in this context, offering a means to verify human authorship and maintain the integrity of academic and literary works against the backdrop of AI advancements.

### III. Proposed Framework: Writer's Integrity

The "Writer's Integrity" framework is designed to authenticate the human origin of textual content, distinguishing it from AI-generated text. This framework is particularly crucial for academic, research, and publishing sectors where the authenticity of human intellectual effort is paramount. Unlike existing tools that focus on analyzing textual patterns to detect AI-generated content, the "Writer's Integrity" framework evaluates the writing process, including typing speed, frequency of edits, and the use of pasted text, to ascertain human authorship.

The "Writer's Integrity" framework operates through a series of processes designed to authenticate the human origin of textual content. Its approach diverges from conventional AI detection tools by focusing on the dynamics of the writing process rather than textual analysis. Figure 1 shows a breakdown of its core processes:

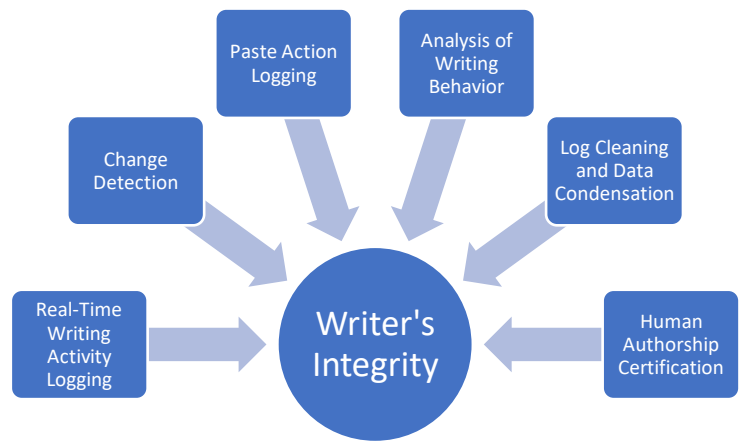

**Fig. 1** Writer's Integrity Framework Processes

1. Real-Time Writing Activity Logging:

The framework initiates by monitoring and logging real-time writing activities in a designated text area. This includes every keystroke, edit (additions and deletions), and paste action performed by the writer.

2. Change Detection:

As the writer composes or modifies text, the framework detects and logs changes between the previous and current states of the text. This includes identifying new words added, existing words removed, or alterations to the text structure.

3. Paste Action Logging:

Special attention is given to text pasted into the document. The framework logs the occurrence of paste actions separately, including the length of the pasted text and its position within the document.

4. Analysis of Writing Behavior:

The collected data is then analyzed to calculate key metrics that characterize human writing behavior, such as typing speed, frequency of edits, and the ratio of pasted text to overall content.

5. Log Cleaning and Data Condensation:

To optimize storage and improve data readability, the framework processes the raw log data, condensing it by removing redundant information and summarizing the writing activity.

6. Human Authorship Certification:

Based on the analyzed data, the framework generates a "Writer's Integrity Certificate" for the document. This certificate provides a summary of the writing behavior metrics, attesting to the human origin of the content.

### A. Metrics

The operational metrics of the framework are defined mathematically as follows:

**Typing Speed (TS):**

$$TS = \frac{\text{Total Typed Characters}}{\text{Total Writing Time}} * 60 \quad (1)$$

Measured in characters per minute, this metric reflects the average speed at which a user types.

**Edit Frequency (EF):**

$$EF = \frac{\text{Total Edits}}{\text{Total Writing Time}} \quad (2)$$

This ratio indicates how frequently the text is edited, embodying the iterative nature of human writing.

**Paste Ratio (PR):**

$$PR = \frac{\text{Total Characters Pasted}}{\text{Total Characters in Document}} \quad (3)$$

Expressed as a percentage, this metric quantifies the proportion of the text that was pasted rather than typed.

**Average Changes per Word (ACW):**

$$ACW = \frac{\text{Total Edits}}{\text{Total Number of Words}} \quad (4)$$

Reflects the average number of edits applied to each word, indicating the depth of revision and thought process.

### B. Pseudocode for Writer's Integrity Framework Functionality

This pseudocode outlines the core functionalities of the "Writer's Integrity" framework, focusing on logging writing activities, detecting changes, cleaning log data, and analyzing writing behavior to identify human-generated texts.

**Logging Writing Activity**

```
Initialize previousText with the current text in the text input field
Initialize typingStartTime with the current date and time
Initialize totalTypedCharacters to 0
When a user types in the text input field:
    Update currentText with the new text value
    If the input is not from pasting:
        Call detectChanges with previousText and currentText
        If changes are detected:
            Call logChanges with the detected changes
    Else if the input is from pasting:
        Call logPastedText with the pasted text
    Update previousText with currentText
    Update totalTypedCharacters based on the number of characters typed
    Update typing speed using current time, typingStartTime, and totalTypedCharacters
```

**Detecting Changes**

Function detectChanges takes oldText and newText as inputs:

    Split oldText and newText into arrays of words, oldWords and newWords respectively

    Build an LCS (Longest Common Subsequence) matrix for oldWords and newWords

    Initialize changes as an empty string

    Using the LCS matrix, backtrack to find differences:

        If words are the same, move diagonally and continue

        If words differ, record as "added" or "removed" in changes

    Return changes

**Logging Changes and Pasted Text**

Function logChanges takes changes as input:

    Record changes with a timestamp to a log

Function logPastedText takes pastedText as input:

    Find the position of the pastedText in the currentText

    Log the pastedText with its position and a timestamp

**Updating Typing Speed**

Function updateTypingSpeed takes charCount, startTime, and totalChars as inputs:

    Calculate the time elapsed since startTime

    Update totalChars with charCount

    Calculate typing speed as totalChars divided by time elapsed

    Display or record the typing speed

**Cleaning Log Data**

```
Function CleanLog takes logs as input:
    Extract "Pasted" log entries and "Added/Removed" entries from logs
    For each "Pasted" entry, add to pastedRecords
    For each "Added/Removed" entry, process and add to finalRecords
    Combine pastedRecords and finalRecords into a cleaned log
    Return the cleaned log
```

**Analyzing Log Data**

```
Function AnalyzeLog takes logData as input:
    Split logData into individual entries
    For each entry, extract relevant data (e.g., pastes, edits)
    Calculate metrics such as number of pastes, total pasted words, total typed words, and average changes per word
    Calculate typing speed and paste to type ratio
    Output or display the analysis results
```

This structured approach allows the "Writer's Integrity" framework to capture and analyze the intricacies of human writing behavior, differentiating it from AI-generated content through detailed logging and analysis of the writing process.

## VI. Empirical Experimental Results of the Writer's Integrity Framework

The "Writer's Integrity" framework has been developed using ASP.Net and VB.net, with SQL Server managing the database operations. This section delves into the empirical experimental results derived from the deployment and testing of the framework, illustrating its effectiveness in ensuring the integrity of human-generated text.

### A. System Overview

The framework operates through a user-friendly interface that begins with a login screen, ensuring secure access to users' documents. Once logged in, users are presented with their saved documents, offering options to edit existing documents or create new ones. The primary functionality kicks in as the user starts typing: every keystroke, edit, and paste action is meticulously logged. Upon saving a document, the logs are cleaned to optimize storage efficiency, and a unique certificate ID is generated for the document.

**B. Key Features and Results**

1. Document Management: Users can seamlessly manage their documents, with each action from creation to modification being intuitively integrated within the system's user interface.

2. Change Logging: The real-time logging of text changes and paste actions serves as the backbone of the framework. This feature not only captures the essence of the writing process but also aids in distinguishing human authorship from AI-generated content.

3. Log Cleaning and Certificate Generation: The log cleaning process, which is triggered upon saving a document, efficiently condenses the log data without losing the critical information necessary for verification. Each document is assigned a unique certificate ID, facilitating easy sharing and verification by reviewers or supervisors.

4. Certificate Verification: The certificate ID acts as a key to access the document's log, presenting a detailed account of all changes and pasted text. This transparency ensures that reviewers can thoroughly assess the document's authenticity.

**C. Screenshots and Logs**

Screenshots of the system in action and samples of the logs before and after cleaning were instrumental in visualizing the framework's functionality. These empirical evidences underscore the framework's capability to maintain a detailed and verifiable record of the writing process. Figure 2 depicts the initial interface where users authenticate their access, while Figure 3 displays the user's collection of stored documents. Figure 4 presents the interface for text input, where the user's typing activity is recorded. In Figure 5, we observe the detailed log and statistics associated with a document, as identified by its unique certificate ID. Lastly, Figure 6 illustrates an example of a log that includes pasted text and highlights the calculated paste ratio.

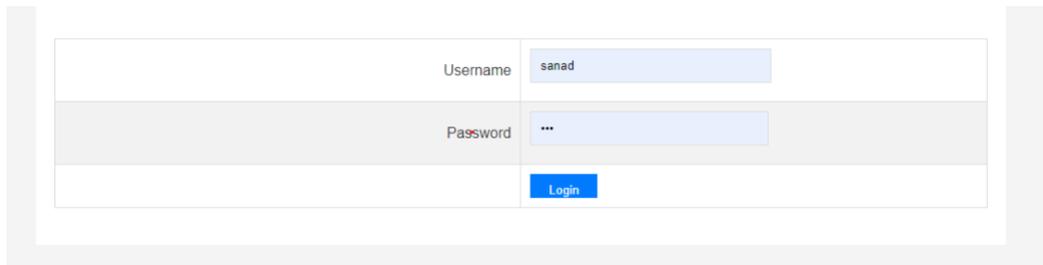

**Fig. 2** User Authentication Screen

**Fig. 3** Document Management Dashboard

**Fig. 4** Text Editing Interface

**Fig. 5** Document Log and Statistical Analysis Display

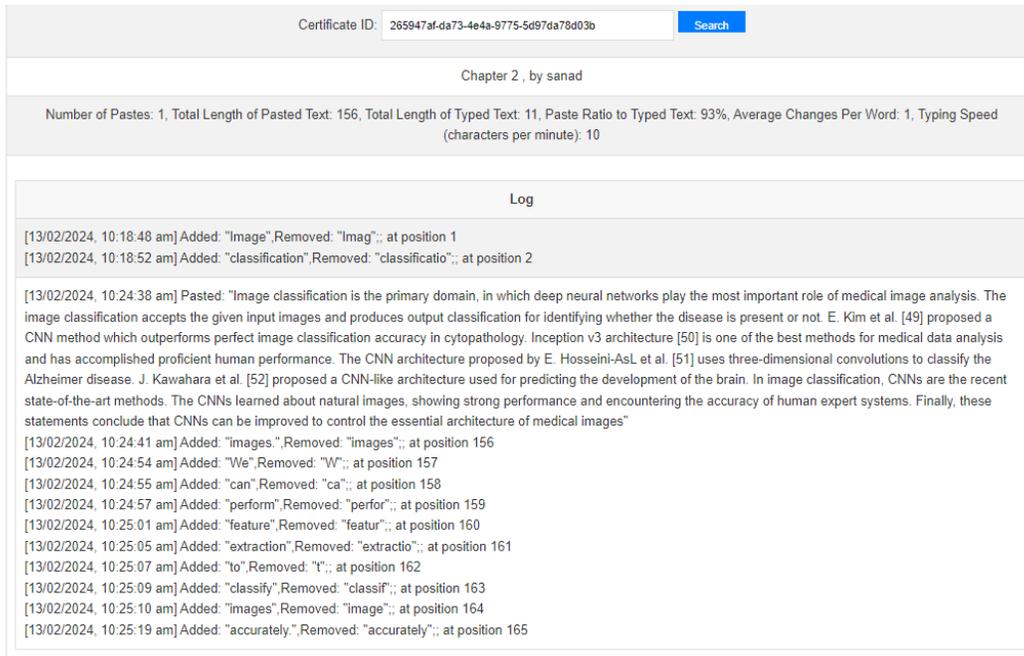

**Fig. 6** Log Excerpt with Pasted Text and Paste Ratio Indicator

### D. The Mechanism of Log Detailing

In the Writer's Integrity Framework, the writing process is meticulously logged to capture every textual change. Each addition and deletion is timestamped and recorded with positional accuracy. For example, as a user composes the phrase "Convolutional Neural Network is a methodology of deep learning that is based on feature extraction," the system logs each incremental input, as shown in table 1.

**Table 1** An example of a log before and after cleaning

| The Original Log | | The Cleaned Log |
|---|---|---|
| [13/02/2024, 10:14:37 am] Added: "C",Removed: "";; at position 1 | [13/02/2024, 10:15:06 am] Added: "",[13/02/2024, 10:15:08 am] Added: "a",Removed: "";; at position 5 | [13/02/2024, 10:14:43 am] Added: "Convolutional",Removed: "Convolutiona";; at position 1 |
| [13/02/2024, 10:14:37 am] Added: "Co",Removed: "C";; at position 1 | [13/02/2024, 10:15:08 am] Added: "",[13/02/2024, 10:15:09 am] Added: "m",Removed: "";; at position 6 | [13/02/2024, 10:14:47 am] Added: "Neural",Removed: "Neura";; at position 2 |
| [13/02/2024, 10:14:37 am] Added: "Con",Removed: "Co";; at position 1 | [13/02/2024, 10:15:09 am] Added: "me",Removed: "m";; at position 6 | [13/02/2024, 10:15:54 am] Added: "Network",Removed: "Network,";; at position 3 |
| [13/02/2024, 10:14:38 am] Added: "Conv",Removed: "Con";; at position 1 | [13/02/2024, 10:15:10 am] Added: "met",Removed: "me";; at position 6 | [13/02/2024, 10:15:05 am] Added: "is",Removed: "i";; at position 4 |
| [13/02/2024, 10:14:41 am] Added: "Convo",Removed: "Conv";; at position 1 | [13/02/2024, 10:15:10 am] Added: "meth",Removed: "met";; at position 6 | [13/02/2024, 10:15:06 am] Added: "",[13/02/2024, 10:15:08 am] Added: "a",Removed: "";; at position 5 |
| [13/02/2024, 10:14:41 am] Added: "Convol",Removed: "Convo";; at position 1 | [13/02/2024, 10:15:11 am] Added: "metho",Removed: "meth";; at position 6 | [13/02/2024, 10:15:13 am] Added: "",[13/02/2024, 10:15:16 am] Added: "methodology",Removed: "metholdolgy";; at position 6 |
| [13/02/2024, 10:14:42 am] Added: "Convolu",Removed: "Convol";; at position 1 | [13/02/2024, 10:15:11 am] Added: "methol",Removed: "metho";; at position 6 | [13/02/2024, 10:15:19 am] Added: "o",[13/02/2024, 10:15:19 am] Added: "of",Removed: "o";; at position 7 |
| [13/02/2024, 10:14:42 am] Added: "Convolut",Removed: "Convolu";; at position 1 | [13/02/2024, 10:15:12 am] Added: "methold",Removed: "methol";; at position 6 | [13/02/2024, 10:15:21 am] Added: "deep",Removed: "dee";; at position 8 |
| [13/02/2024, 10:14:42 am] Added: "Convoluti",Removed: "Convolut";; at position 1 | [13/02/2024, 10:15:12 am] Added: "methodo",Removed: "methold";; at position 6 | [13/02/2024, 10:15:22 am] Added: "learning",Removed: "learnin";; at position 9 |
| [13/02/2024, 10:14:42 am] Added: "Convolutio",Removed: "Convoluti";; at position 1 | [13/02/2024, 10:15:13 am] Added: "metholdol",Removed: "methodo";; at position 6 | [13/02/2024, 10:15:27 am] Added: "that",Removed: "tha";; at position 10 |
| [13/02/2024, 10:14:43 am] Added: "Convolution",Removed: "Convolutio";; at position 1 | [13/02/2024, 10:15:13 am] Added: "metholdolg",Removed: "metholdol";; at position 6 | [13/02/2024, 10:15:27 am] Added: "i",[13/02/2024, 10:15:27 am] Added: "is",Removed: "i";; at position 11 |
| [13/02/2024, 10:14:43 am] Added: "Convolutiona",Removed: "Convolution";; at position 1 | [13/02/2024, 10:15:13 am] Added: "metholdolgy",Removed: "metholdolg";; at position 6 | [13/02/2024, 10:15:28 am] Added: "based",Removed: "base";; at position 12 |
| [13/02/2024, 10:14:43 am] Added: "Convolutional",Removed: "Convolutiona";; at position 1 | [13/02/2024, 10:15:13 am] Added: "",[13/02/2024, 10:15:16 am] Added: "methodology",Removed: "metholdolgy";; at position 6 | |

[13/02/2024, 10:14:43 am] Added: "",[13/02/2024, 10:14:45 am] Added: "N",Removed: "";; at position 2
[13/02/2024, 10:14:46 am] Added: "Ne",Removed: "N";; at position 2
[13/02/2024, 10:14:46 am] Added: "Neu",Removed: "Ne";; at position 2
[13/02/2024, 10:14:47 am] Added: "Neur",Removed: "Neu";; at position 2
[13/02/2024, 10:14:47 am] Added: "Neura",Removed: "Neur";; at position 2
[13/02/2024, 10:14:47 am] Added: "Neural",Removed: "Neura";; at position 2
[13/02/2024, 10:14:47 am] Added: "",[13/02/2024, 10:14:48 am] Added: "N",Removed: "";; at position 3
[13/02/2024, 10:14:49 am] Added: "Ne",Removed: "N";; at position 3
[13/02/2024, 10:14:49 am] Added: "Net",Removed: "Ne";; at position 3
[13/02/2024, 10:14:49 am] Added: "Netw",Removed: "Net";; at position 3
[13/02/2024, 10:14:50 am] Added: "Netwo",Removed: "Netw";; at position 3
[13/02/2024, 10:14:50 am] Added: "Networ",Removed: "Netwo";; at position 3
[13/02/2024, 10:14:51 am] Added: "Network",Removed: "Networ";; at position 3
[13/02/2024, 10:14:51 am] Added: "Networks",Removed: "Network";; at position 3
[13/02/2024, 10:14:52 am] Added: "Networks,",Removed: "Networks";; at position 3
[13/02/2024, 10:14:53 am] Added: "",[13/02/2024, 10:14:53 am] Added: "a",Removed: "";; at position 4
[13/02/2024, 10:14:54 am] Added: "ar",Removed: "a";; at position 4
[13/02/2024, 10:14:54 am] Added: "are",Removed: "ar";; at position 4
[13/02/2024, 10:14:54 am] Added: "",[13/02/2024, 10:14:59 am] Added: "a",Removed: "";; at position 5
[13/02/2024, 10:15:00 am] Added: "",[13/02/2024, 10:15:01 am] Removed: "";; at position 6
[13/02/2024, 10:15:01 am] Added: "",Removed: "a";; at position 5
[13/02/2024, 10:15:02 am] Removed: "";; at position 5
[13/02/2024, 10:15:02 am] Added: "ar",Removed: "are";; at position 4
[13/02/2024, 10:15:02 am] Added: "a",Removed: "ar";; at position 4
[13/02/2024, 10:15:03 am] Added: "",Removed: "a";; at position 4
[13/02/2024, 10:15:03 am] Removed: "";; at position 4
[13/02/2024, 10:15:03 am] Added: "Networks",Removed: "Networks,";; at position 3
[13/02/2024, 10:15:03 am] Added: "Network",Removed: "Networks";; at position 3
[13/02/2024, 10:15:04 am] Added: "Network,",Removed: "Network";; at position 3
[13/02/2024, 10:15:04 am] Added: "",[13/02/2024, 10:15:05 am] Added: "i",Removed: "";; at position 4
[13/02/2024, 10:15:05 am] Added: "is",Removed: "i";; at position 4
[13/02/2024, 10:15:19 am] Added: "o",[13/02/2024, 10:15:19 am] Added: "of",Removed: "o";; at position 7
[13/02/2024, 10:15:20 am] Added: "d",[13/02/2024, 10:15:20 am] Added: "de",Removed: "d";; at position 8
[13/02/2024, 10:15:20 am] Added: "dee",Removed: "de";; at position 8
[13/02/2024, 10:15:21 am] Added: "deep",Removed: "dee";; at position 8
[13/02/2024, 10:15:21 am] Added: "l",[13/02/2024, 10:15:21 am] Added: "le",Removed: "l";; at position 9
[13/02/2024, 10:15:21 am] Added: "lea",Removed: "le";; at position 9
[13/02/2024, 10:15:22 am] Added: "lear",Removed: "lea";; at position 9
[13/02/2024, 10:15:22 am] Added: "learn",Removed: "lear";; at position 9
[13/02/2024, 10:15:22 am] Added: "learni",Removed: "learn";; at position 9
[13/02/2024, 10:15:22 am] Added: "learnin",Removed: "learni";; at position 9
[13/02/2024, 10:15:22 am] Added: "learning",Removed: "learnin";; at position 9
[13/02/2024, 10:15:26 am] Added: "t",[13/02/2024, 10:15:26 am] Added: "th",Removed: "t";; at position 10
[13/02/2024, 10:15:27 am] Added: "tha",Removed: "th";; at position 10
[13/02/2024, 10:15:27 am] Added: "that",Removed: "tha";; at position 10
[13/02/2024, 10:15:27 am] Added: "i",[13/02/2024, 10:15:27 am] Added: "is",Removed: "i";; at position 11
[13/02/2024, 10:15:28 am] Added: "b",[13/02/2024, 10:15:28 am] Added: "ba",Removed: "b";; at position 12
[13/02/2024, 10:15:28 am] Added: "bas",Removed: "ba";; at position 12
[13/02/2024, 10:15:28 am] Added: "base",Removed: "bas";; at position 12
[13/02/2024, 10:15:28 am] Added: "based",Removed: "base";; at position 12
[13/02/2024, 10:15:29 am] Added: "o",[13/02/2024, 10:15:29 am] Added: "on",Removed: "o";; at position 13
[13/02/2024, 10:15:32 am] Added: "f",[13/02/2024, 10:15:32 am] Added: "fe",Removed: "f";; at position 14
[13/02/2024, 10:15:32 am] Added: "fea",Removed: "fe";; at position 14
[13/02/2024, 10:15:33 am] Added: "feat",Removed: "fea";; at position 14
[13/02/2024, 10:15:33 am] Added: "featu",Removed: "feat";; at position 14
[13/02/2024, 10:15:33 am] Added: "featur",Removed: "featu";; at position 14
[13/02/2024, 10:15:33 am] Added: "feature",Removed: "featur";; at position 14
[13/02/2024, 10:15:33 am] Added: "e",[13/02/2024, 10:15:34 am] Added: "ex",Removed: "e";; at position 15
[13/02/2024, 10:15:37 am] Added: "extraction",Removed: "ex";; at position 15
[13/02/2024, 10:15:48 am] Added: "extraction.",Removed: "extraction";; at position 15
[13/02/2024, 10:15:54 am] Added: "Network",Removed: "Network,";; at position 3

### E. The Original Log

The original log serves as a comprehensive record, tracking the user's every action. It begins with the initial characters as they are typed, such as "C," "Co," "Con," and continues as more of the word "Convolutional" is formed. Each logged entry includes the exact time of the edit and the specific change made character additions and deletions at precise positions within the text. This level of detail is crucial for supervisors or reviewers who need to verify the authenticity of the writing process. It ensures that the human intellect and effort behind the text can be observed and validated.

### F. Log Cleaning Process

To maintain an efficient and readable log, redundant entries that capture the step-by-step typing process are consolidated to reflect only significant changes, such as the completion of words or phrases. The cleaning process highlights the final version of words or phrases, as they appear in the completed text, and associates them with the corresponding timestamp and change history. For instance, multiple entries recording the sequential typing of "Convolutional" are condensed into a single entry in the cleaned log, showcasing the word as it is fully typed along with the last removal that occurred before its completion.

### G. The Cleaned Log

The cleaned log presents a streamlined view, it removes the minute-by-minute keystroke details, instead of providing a clear snapshot of the significant changes that occurred during the writing session. This cleaned version is not only easier to read but also focuses on the end result of the writing process, which is more relevant for authenticity verification. Highlighting in the cleaned log allows reviewers to quickly discern the final text and the key changes, enabling them to assess the integrity of the document efficiently.

By implementing such detailed logging and cleaning mechanisms, the Writer's Integrity Framework ensures that the human-generated content can be effectively differentiated from AI-generated text. The logs provide transparency and accountability, which are essential in settings where authorship and original thought are of the utmost importance.

## VI. Implementation Considerations for the Writer's Integrity Framework

### A. Ease of Use and Privacy Concerns

The implementation of the Writer's Integrity Framework necessitates a delicate balance between ease of use and privacy. While some students and writers might initially perceive the system as intrusive, it is imperative to understand the context and necessity of such a framework. In academia and professional writing, particularly where Ph.D. dissertations, published research, or career advancements are at stake, the authenticity of the text is paramount. Monetary compensation or academic credentials based on the text further necessitate this verification.

Users should be reassured that, much like proctoring systems used during online examinations, the Writer's Integrity Framework is a tool designed to uphold academic and professional standards. If the text is original and the authors have nothing to conceal, the use of this system should be viewed similarly to any other academic integrity practice. It is a safeguard, not a surveillance measure, ensuring that the credit and recognition received are rightfully earned.

### B. Drafting System Focus

This framework is tailored as a drafting system, specifically designed for the initial creation and development of text. It is not intended to replace comprehensive word-processing software. Users are encouraged to utilize the Writer's Integrity Framework for drafting their work, benefiting from the robust logging and authenticity verification processes. Subsequently, they can transfer their drafts into their preferred word-processing programs to finalize formatting, insert figures, and add tables. Adopting this specialized approach allows implementers to concentrate on refining the core functionalities that differentiate human-authored content from AI-generated text. This focus ensures the framework's resources are dedicated to enhancing the key features that support its primary purpose—authenticating the integrity of written work.

### C. Privacy and Security

Given the sensitive nature of the documents handled by the Writer's Integrity Framework—ranging from confidential research to doctoral dissertations—the system is engineered with stringent security measures. The privacy and protection of intellectual property are of utmost importance. As such, the framework must incorporate advanced security protocols to safeguard against unauthorized access and data breaches. Security measures, including robust encryption, secure authentication mechanisms, and regular security audits, are vital to protect the integrity of the text and the privacy of the users. The implementation of these measures ensures that while the framework serves its function in verifying authorship, it also maintains the confidentiality and security that users rightfully expect for their scholarly and professional work.

### D. Business Model for the Writer's Integrity Framework

The Writer's Integrity Framework presents a unique opportunity for tech companies to offer a valuable service to universities, publishing companies, and individual writers. The framework's ability to verify the human origin of text is a critical asset in an era where the distinction between human and AI-generated content is increasingly blurred. Below is an outline of a potential business model that companies can adopt to monetize this framework:

### E. Licensing to Academic Institutions

Universities are prime candidates for implementing this system to preserve academic integrity. Tech companies can offer the framework as a licensed product, allowing institutions to integrate it into their existing digital infrastructure. A tiered licensing model can be established, depending on the size of the institution and usage volume, ensuring scalability and accessibility for universities with varying budgets and requirements.

### F. Partnerships with Publishing Entities

Publishers, research journals, and press agencies are stakeholders in ensuring the authenticity of published content. A tech company can partner with these organizations, offering licenses to use the framework to vet submissions efficiently. Such partnerships could involve a flat fee for access to the system or a pay-per-use model, providing flexibility for publishers with fluctuating volumes of content to authenticate.

### G. Subscriptions for Individual Writers

Independent writers seeking to certify the originality of their work can benefit from subscribing to the Writer's Integrity Framework. Subscription tiers can be established based on usage, such as the number of texts to be authenticated or a specific period. A premium model can offer additional features like advanced analytics on writing habits, priority support, and storage options for documented logs and certificates.

### H. Verified Profiles for Writers

Tech companies can create a platform where writers can have verified profiles, akin to social media verifications. These profiles would display authenticated certificates from the Writer's Integrity Framework, showcasing the writer's commitment to originality. Verified profiles could include metrics reflecting the individual's typing habits and other behavioral analytics that reinforce their status as human authors. Access to these profiles could be provided to publishing companies, serving as a credibility benchmark for writers. This could be monetized by charging writers a fee for verification or offering subscription-based access to publishers.

### I. Data Privacy Services

For institutions and individuals concerned with data privacy, tech companies can offer additional security services. These might include on-premises implementations, encrypted data storage, or private cloud services. These specialized privacy services can be offered as part of a premium package or as add-ons to the standard licensing or subscription model.

### J. Ancillary Services

Tech companies can also explore ancillary services such as training for university staff, workshops for writers on maintaining integrity, and consultation for publishing companies on integrating the framework into their workflow. These services can be charged separately, providing a supplementary revenue stream while adding value to the core offerings of the framework.

By leveraging these monetization strategies, tech companies can build a profitable business while providing a critical service that upholds the integrity of written content across various industries and sectors.

## V. Conclusion

In conclusion, the "Writer's Integrity" framework heralds a new era in the verification of human-generated text, crucial for academic, research, and publishing sectors. The system's innovative approach, focusing on the writing process's dynamics rather than solely on the final product, offers a robust solution to the challenges posed by AI-generated content. By analyzing metrics such as typing speed, change logs, and the ratio of pasted to original text, the framework can effectively discern between human and AI authorship, providing a level of assurance that traditional AI detection tools have struggled to achieve. The practical implementation of this framework has been carefully considered, with attention given to user experience and privacy concerns. By positioning the system as a drafting tool and ensuring high-security standards, the framework respects the privacy of its users while providing a secure environment for the creation of original works. From a business perspective, the "Writer's Integrity" framework presents a viable model for companies aiming to implement this technology. By offering various licensing and subscription options to universities, publishing houses, and individual authors, the system promises to open up new revenue streams while bolstering the integrity of written works. As we move forward, the adoption of "Writer's Integrity" could become a standard in industries where the authenticity of human intellect is of paramount importance. This would not only safeguard the value of human creativity and critical thinking but also maintain the credibility and authenticity of human contributions in an increasingly AI-assimilated world.

Declarations

Availability of data and materials: This framework is open source and available here,

https://github.com/sanadv/Integrity.github.io

Funding: Not applicable.

Acknowledgements: Not applicable.